\documentclass[9pt,twocolumn,twoside, lineno]{pnas-new}

\templatetype{pnasresearcharticle} 

\newcommand{\B}{\ensuremath{\mathcal{B}}}

\begin{document}

\title{
Free Information Disrupts Even Bayesian Crowds
}

\author[a,1]{Jonas Stein}
\author[b,1]{Shannon Cruz}
\author[c,1]{Davide Grossi}
\author[d,1,$\ast$]{Martina Testori}

\affil[a]{Faculty of Behavioural and Social Sciences, University of Groningen, The Netherlands.}
\affil[b]{Department of Communication Arts and Sciences, The Pennsylvania State University, USA}
\affil[c]{Bernoulli Institute for Mathematics, Computer Science and Artificial Intelligence, University of Groningen; Amsterdam Center for Law \& Economics, University of Amsterdam; Institute for Logic, Language and Computation, University of Amsterdam, The Netherlands.
}
\affil[d]{Greenwich Business School, University of Greenwich, Old Royal Naval College, UK}

% Please give the surname of the lead author for the running footer --> first shared coauthorship
%\leadauthor{}

\significancestatement{
A foundational assumption in the design of digital communication platforms is that information sharing is always beneficial. Using an agent-based model of truth-seeking and cooperative agents with perfect information-processing abilities, we show that unconstrained information exchange can reduce the accuracy of group beliefs, particularly in socially homophilous settings. Because these negative effects emerge even in crowds of agents that process information perfectly and cooperatively,
a plausible risk exists of them being 
 even more severe in less ideal, real-world contexts.
Our findings challenge the common design principle that information should always “be free”, in the sense of being unlimitedly shareable, and suggest that constraining the amount of information shared — rather than maximizing it — may actually improve collective epistemic outcomes.}

\authorcontributions{\textsuperscript{1}J.S., S.C., D.G., and M.T. contributed equally to this work.}
%\authordeclaration{The authors declare no competing interests.}
%\correspondingauthor{\textsuperscript{$\ast$}To whom correspondence should be addressed. E-mail: testori.martina@gmail.com}

\keywords{Wisdom of crowds $|$ Homophily $|$ Epistemic inequality $|$ Social learning $|$ Bayesian agent based model}

\begin{abstract}
A core tenet underpinning the conception of contemporary information networks, such as social media platforms, is that users should not be constrained in the amount of information they can freely and willingly exchange with one another about a given topic. By means of a computational agent-based model, we show how even in groups of 
truth-seeking and cooperative agents with perfect information-processing abilities,
unconstrained information exchange may lead to detrimental effects on the correctness of the group's beliefs. If unconstrained information exchange can be detrimental even 
among such idealized agents, it is prudent to assume it can also be so in practice. We therefore argue that constraints on information flow should be carefully considered in the design of communication networks with substantial societal impact, such as social media platforms.
\end{abstract}

\dates{
This is a preprint of an article published in Proceedings of the National Academy of Sciences of the United States of America (PNAS). The final authenticated version, with supplementary materials, is available online at: \fbox{\url{https://doi.org/10.1073/pnas.2518472123}}
%This manuscript was compiled on \today
}

%\doi{\url{www.pnas.org/cgi/doi/10.1073/pnas.XXXXXXXXXX}}

\maketitle
\thispagestyle{firststyle}
\ifthenelse{\boolean{shortarticle}}{\ifthenelse{\boolean{singlecolumn}}{\abscontentformatted}{\abscontent}}{}

\dropcap{A} cornerstone view in the contemporary conception of digital communication is that no piece of information between two individuals is ever one piece too many, when individuals are willing to communicate with one another. Crucially, this idea still pervades the narrative projected by social media platforms around their societal role, rooted in the digital activism mantra that ``information wants to be free'' \cite{denning1996concerning}. Even arguments in favor of, or opposed to, specific technical design features of social media platforms, such as full-text search of posts, quote-posts, or even fact-checking, are often framed with respect to the conviction that information generation by users should be easy and information provision to users should be seamless.
This idea extends beyond social media, permeating broader discussions around openness in the digital age. Exchanges between practitioners  in education \cite{UNESCO_OER_2025}, patients and clinicians in healthcare \cite{wolff2017inviting}, or researchers collaborating across institutions \cite{Arentoft2022_Agreement} are frequently described in similar terms: The more information is shared among individuals, the more epistemically effective the interaction becomes. In these domains, the free flow of information is seen as essential for fostering mutual learning, improving patient treatment plans, and enhancing the quality and impact of research, thereby reinforcing its inherent value.

In this paper, we scrutinize the above tenet from a wisdom-of-crowds perspective \cite{grofman1983thirteen}. In their simplest form, these settings involve a group of agents, each endowed with private information about the value of a binary variable of interest. The group then needs to aggregate the members' private information and make a collective decision about the correct value of the (binary) variable. A substantial body of work using both experimental and computational techniques (see \cite{centola2022network} for a recent overview) has already shown that information sharing networks strongly influence the epistemic performance of groups in such settings and that, crucially, different network topologies induce different epistemic effects.  Within such a context, our paper analyzes epistemic performance of groups as a function of two parameters of peer interaction that abstract away from the explicit consideration of network topology: homophily and information-exchange capacity. Specifically, we minimally enrich the classical wisdom-of-crowds setting by allowing our agents to reveal their private information to one another in bilateral encounters. In doing so, we follow recent work \cite{hahn2019communication,hahn2020truth,becker2017network,lorenz2011social} but enrich it in one specific dimension by manipulating how much private information can be revealed in such encounters.

This simple feature allows us to investigate the effects of various levels of information-sharing capacity and homophily on the epistemic performance of crowds. Importantly, we investigate such information-sharing effects under the most ideal information processing and sharing conditions during encounters: a crowd consisting of perfect, i.e., Bayesian, information-processing agents, who communicate truthfully and fully cooperatively. 
At the same time, we place these agents in interaction conditions that capture a pervasive feature of many real-world information networks: homophily.
The choice of these specific model features is
methodologically motivated (we do, however, relax them to ascertain robustness of our findings, see SI Section 3): If unlimited information exchange has negative impact under even the most ideal assumptions about individuals, 
it should be questioned altogether as a foundational design principle for information-sharing networks. 

Through agent-based simulations, we observe that the ability to exchange more information negatively affects the decision-making performance of the group 
when information is shared in the context of homophilous encounters, 
even under such ideal individual information processing and sharing conditions. 
This largely aligns with previous findings \cite{centola2022network}, while focusing on the combined effects of homophily and information-sharing capacity.
Our contribution should be interpreted as a computationally-aided thought experiment, showing how the ``information wants to be free'' principle in the design of digital communication networks should be subject to qualification:
being constrained in how much information can be shared may actually improve collective decision making when homophily is high.

\begin{figure}[t]
\centering
\includegraphics[width=\linewidth]{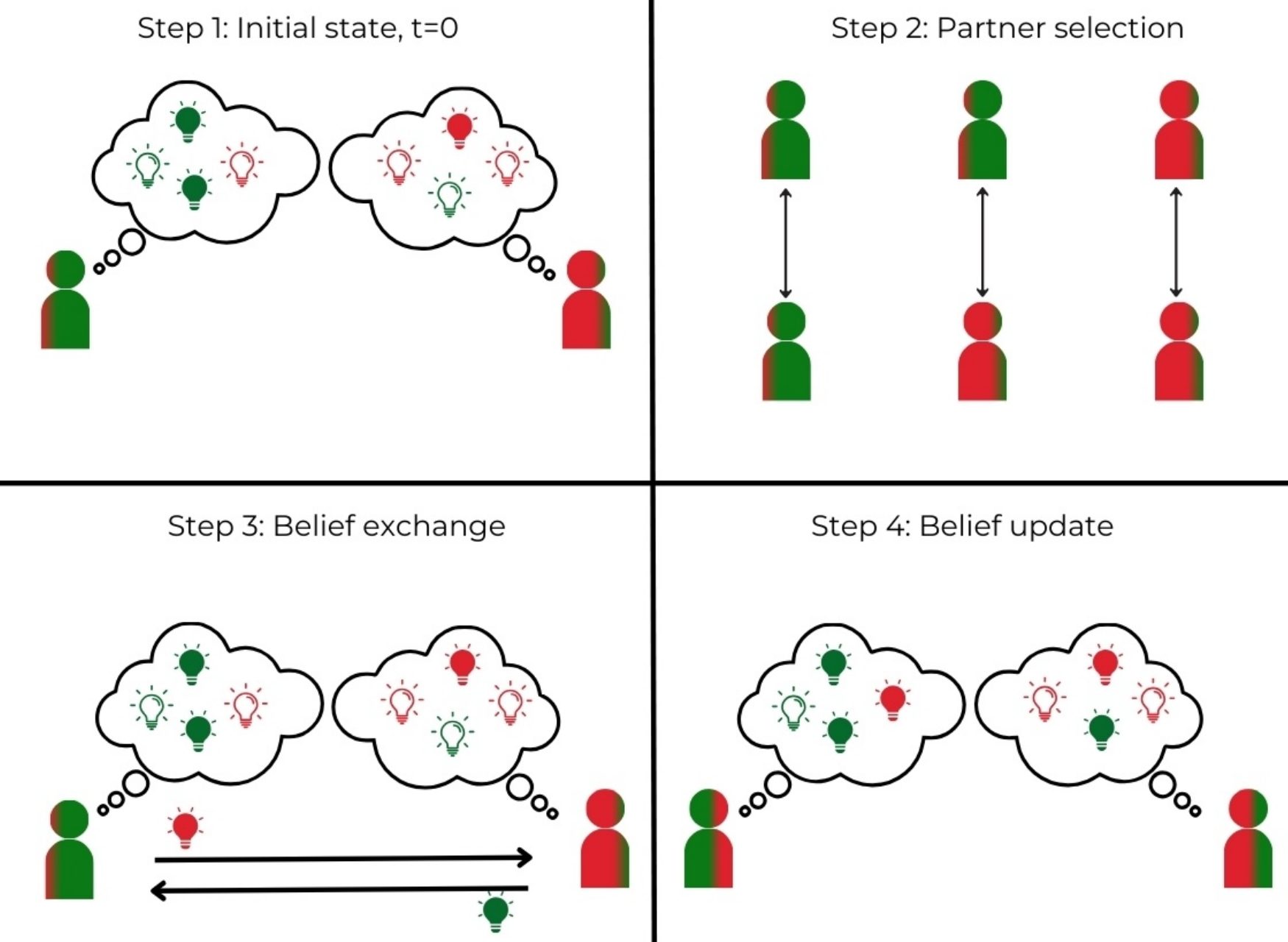}
\caption{Each agent is represented by a colored figure, with the color indicating their belief: Green-leaning agents are more confident that the true state is {\em A}, while red-leaning agents are more confident in the false state {\em B}.
Step 1: Initial state ({\em t} = 0).
Each agent is endowed with a private piece of evidence about the state of the world. Each piece of evidence (green lightbulb for {\em A}, red lightbulb for {\em B}) comes with a known quality — the probability that the signal is accurate (full/empty lightbulb for high/low quality). Agents use Bayes’ rule to form an initial belief based on this single observation. Some agents start with a correct belief (favoring {\em A}), others with an incorrect one (favoring {\em B}) due to randomness in their initial evidence.
Step 2: Partner selection.
At each time step, agents are stochastically matched with another agent to exchange information. The likelihood of being paired increases with belief similarity, a dynamic controlled by homophily: When homophily is high, agents are more likely to interact with others who hold beliefs similar to their own.
Step 3: Belief exchange.
Agents exchange up to {\em k} pieces of evidence, determined by their communication capacity. Each agent selects their best available evidence — i.e., the most accurate signals — and shares them in proportion to their current belief. For example, an agent strongly favoring {\em A} will mostly share high-quality signals supporting {\em A}, but may also share some evidence supporting {\em B}. Note: Exchanges occur in one direction, meaning that one agent might not share information with the same agent they received information from.
Step 4: Belief update.
After the exchange, each agent updates their belief by applying Bayes’ rule to the new evidence received. Over time, these updates can lead agents to converge toward the true state {\em A}, though this depends on both who they interact with (influenced by homophily) and how much information they can exchange (communication capacity).}
\label{fig:Info}
\end{figure}

\paragraph{Model.} We model crowds using an agent-based model \cite{flache2021agent} and investigate the interplay of two key parameters of the model: {\em communication capacity} (i.e., the number of pieces of evidence that agents are able to exchange in each interaction) and the level of {\em homophily} that guides agents' interactions (see Figure \ref{fig:Info}). Homophily, the tendency to interact with similar others, is a strong force in humans \cite{mcpherson2001birds} and one that deeply shapes interaction in social and digital information networks, especially when algorithmically curated. Algorithmic curation (i.e., the computational process that filters the information to which users are exposed) is a pervasive design feature of contemporary social media and one that has been repeatedly linked to homophilous information exchange \cite{bakshy2015exposure,boutyline2017social,santos2021link}. 

Our simulation experiments assume a crowd of Bayesian agents seeking to learn the true state of the world. Each agent is endowed with one piece of evidence that either points towards the true state $A$ or the false state $B$ of the world (see Materials and Methods and SI for more details). 
Agents in the model form beliefs by applying the Bayes' rule on the pieces of evidence they have access to. They influence each other by exchanging pieces of evidence. Each piece of evidence consists of two elements: a signal (pointing to either state $A$ or state $B$) and the quality of that signal (i.e., the probability that the signal is correct). This setup can be illustrated by thinking of the agents as carrying out experiments about the state of the world using imperfect experimental equipment of known accuracy. A piece of evidence is thus the outcome of an observation together with information about the accuracy of the equipment that generated the observation. It is important to stress that equipment with higher accuracy can still occasionally return wrong observations, albeit less frequently than equipment with lower accuracy, and, vice versa, equipment with low accuracy can still provide correct observations. As a consequence, accurate evidence may still point to the incorrect state of the world, and inaccurate evidence to the correct one. This richer model of what constitutes information
sets our model apart from most existing literature on information dynamics in groups (see Discussion section).

At model initialization, agents observe their own pieces of evidence. Even though we assume that these observations are more likely to be correct than not, a minority of agents will typically hold incorrect beliefs because their imperfect equipment returns the wrong signal by chance. After initialization, agents seek information about others' evidence. The level of homophily present in the group affects how likely these exchanges are to occur between agents with similar beliefs about the state of the world. Furthermore, the communication capacity determines the number of pieces of evidence that can be exchanged in each interaction.
Importantly, evidence exchange is cooperative: Agents share only their highest-quality evidence and select evidence in favor of $A$ or $B$ with probability proportional to their current belief. In other words, agents try as much as possible, within the given communication constraints, to convey their best available evidence. Agents interact until a stable state is reached at which no further belief change occurs.  
Despite the above idealizations on the information processing ability and cooperation of agents, homophily drives agents with incorrect beliefs to interact with agents that hold similarly incorrect beliefs. And if no constraints are in place on how much information can be exchanged during such encounters, agents with incorrect beliefs will necessarily exchange more evidence that points to the incorrect state of the world. This is, in essence, the phenomenon that our experiments set out to scope, quantify, and analyze.

\paragraph{Focus and metrics: effects of information-exchange capacity.} Our main focus is on the group of agents whose initial observations point towards the wrong state, which we refer to as Group $B$ (as opposed to Group $A$). We are interested in assessing the extent to which, under the above ideal conditions, this group of `unlucky' observers is able to improve the quality of their beliefs through interaction, that is, what we refer to as {\em epistemic gain}. Intuitively, the larger the capacity individuals have to communicate evidence with each other, the more likely it is that agents with incorrect initial beliefs will shift toward the correct belief. We will see, however, that this is not necessarily the case. Limiting communication capacity, and therefore the amount of evidence that can be exchanged in each interaction, serves the epistemic gain of the disadvantaged group better under conditions of high homophily. 

\section*{Results}

\begin{figure}[t]
\centering
\includegraphics[width=\linewidth]{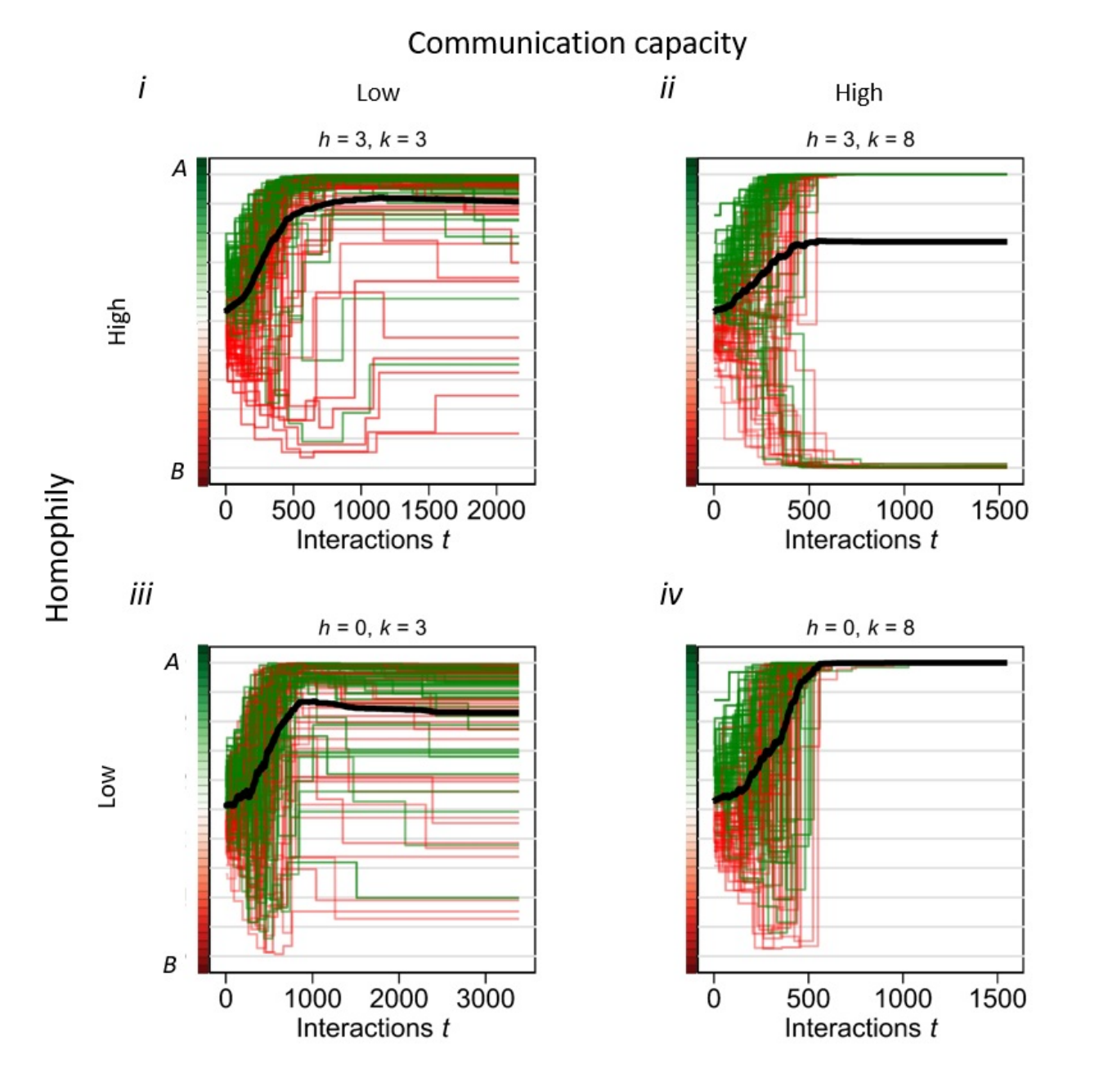}
\caption{Belief evolution over interactions by homophily $h$ and communication capacity $k$. \textbf{Panel i \& iii}: Low communication capacity results in moderate belief diversity and epistemic performance, regardless of the level of homophily. \textbf{Panel ii:} High communication capacity and high homophily result in belief polarization and low epistemic performance. \textbf{Panel iv:} Epistemic performance is maximal at high capacity and no homophily. Green (red) lines represent agents with an initial belief in favor of $A$ ($B$).}
\label{fig:1}
\end{figure}

Our simulation experiments assume a crowd of $N = 100$ agents interacting according to the above model (see Materials and Methods and SI for more details). Even sample runs of the model illustrate that effects of communication capacity are contingent upon homophily (see Fig \ref{fig:1}). Crucially, beliefs only converge around the true state of the world ($A$) when communication capacity is high but homophily is low (Fig \ref{fig:1}, panel iv). In this situation, all agents exchange all relevant information, thereby carrying agents towards the correct belief, regardless of their initial belief. 
Epistemic performance of the population (i.e., how close the average belief is to the correct belief that $A$ is the true state) is lowest in the high-homophily, high-capacity regime (Fig \ref{fig:1}, panel ii), exhibiting a population split into polarized groups at opposite ends of the belief spectrum. Low levels of communication capacity result in moderate epistemic performance, independent of the level of homophily (Fig \ref{fig:1}, panels i,iii).

\begin{figure*}[h]
\centering
\rotatebox{270}{\includegraphics[scale=0.5]{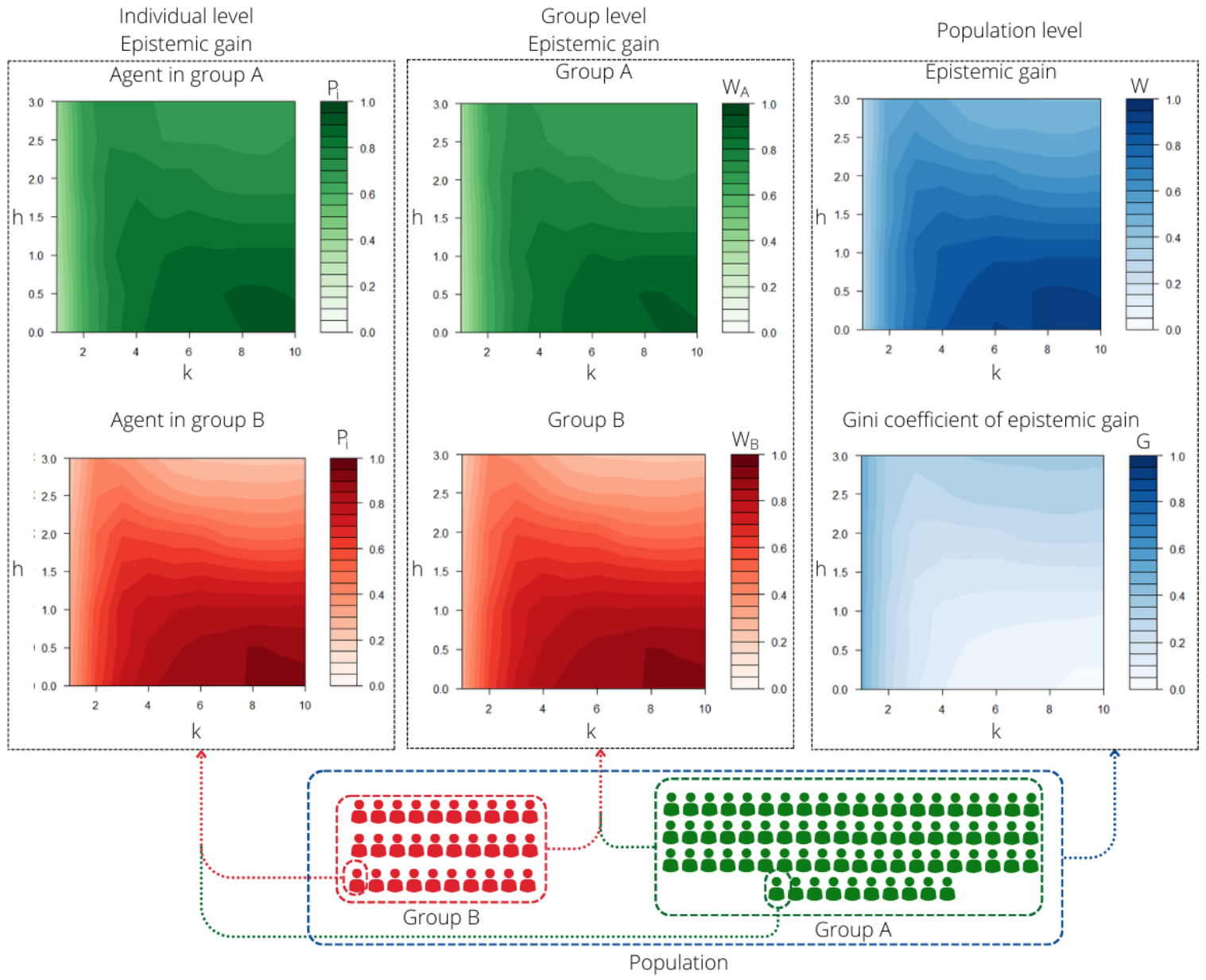}}
\caption{Overview of a) agents' A and B epistemic gain ($\rho_i$; first column); b) groups' A and B epistemic gain ($W_A, W_B$; middle column); c) population epistemic gain ($W$; top panel, last column); d) inequality across the population in the epistemic gain ($G$; bottom panel, last column), as a function of homophily $h$ and communication capacity $k$.}
\label{fig:2}
\end{figure*}

Across model runs, agents' epistemic performance depends not only on homophily and communication capacity, but also on which beliefs agents draw from their private observations at initialization. To examine these effects, we conducted a multilevel analysis with agents (Level 1, $N$ = 100 per run) nested within runs (Level 2, $N$ = 70,000). Individual-level epistemic gain ($\rho_i$) was then regressed on initial beliefs ($B_i$, a level-1 variable), communication capacity ($k$, a standardized level-2 variable), homophily ($h$, a standardized level-2 variable), and the interactions among these predictors. The results (see SI for mathematical details and full tables) suggest that communication capacity by itself has a very limited effect on realized epistemic gain (unstandardized $\beta$ = -0.01). The interaction between communication capacity and homophily ($\beta$ = -0.12), however, suggests that communication capacity tends to exacerbate homophily’s already negative effect on epistemic gain ($\beta$ = -0.48). These effects also differ for agents with different initial beliefs. Agents with initial beliefs that are closer to the correct state of the world (Fig 3, top left) tend to have greater epistemic gain regardless of communication capacity ($\beta$ = 0.63). More importantly, the interaction terms indicate that these agents tend to benefit more from increased communication capacity ($\beta$ = 0.72) and suffer less from the negative effects of homophily ($\beta$ = 0.17) relative to agents with incorrect initial beliefs (Fig 3, bottom left). 

We then probed the group- and population-level outcomes of these dynamics by conducting a series of linear regressions in which we regressed group- and population-level mean epistemic gain and inequality on communication capacity, homophily, and their interaction. At the group level, the results indicate that the mean epistemic gain in both the community of initially correct agents (Group A; $W_A$) and the community of initially incorrect agents (Group $B$; $W_B$) is positively associated with communication capacity, but negatively associated with homophily (see Fig 3, middle column). However, whereas the positive effect of communication capacity outweighs the negative effect of homophily in group $A$ (standardized $\beta$ = .26 vs. -.16), the opposite is true for group $B$ ($\beta$ = .17 vs. -.47). The interaction effects also indicate that, similar to what was observed at the individual level, a higher communication capacity exacerbates the negative effects of homophily more for group $B$ ($\beta$ = -.17) than for group $A$ ($\beta$ = -.12). As a result of these dynamics, the epistemic inequality between the two groups ($W_A-W_B$) tends to get larger as capacity increases ($\beta$ = .11) and as homophily increases ($\beta$ = .65), particularly when both are high simultaneously ($\beta$ = .13).

Because the effects of homophily are particularly harmful to mean epistemic gain for group $B$, the negative effect of homophily ($\beta$ = -.34) also tends to outweigh the positive effect of communication capacity ($\beta$ = .22) on population-level mean epistemic gain ($W$; see Fig 3, top right). Indeed, the interaction between these variables indicates that increased communication capacity exacerbates the negative effects of homophily at the population level ($\beta$ = -.16). The effect of homophily on epistemic inequality, measured as the Gini coefficient of individual epistemic gains ($G$; $\beta$ = .38; see Fig 3, bottom right), also outweighs the equalizing effect of communication capacity ($\beta$ = -.28). When homophily is high, higher capacity also tends to make inequality worse ($\beta$ = .17). 

To be clear, these results focus on how realized epistemic gains compare to the maximum possible epistemic gains. Even in the conditions where gains are lowest, they are rarely zero; information sharing even under the least favorable conditions in our model still results, on average, in some benefit for the population as a whole. The important finding is that under high homophily and high information capacity, epistemic gains are consistently lower {\em relative} to other conditions. In other words, although it is encouraging that epistemic gains occur even under the worst-case scenario in our model, this finding is less positive when viewed against the substantially greater gains that could have been achieved under different information-sharing conditions.

We also conducted model variations and robustness checks (see SI, Section 3) to assess whether the results depend on specific modeling choices. The findings are robust to changes in population size, to shifts in the distribution and number of initial signals assigned to agents, and to modifications of structural features. In particular, results remain mostly consistent when allowing information to be shared simultaneously with multiple agents (one-to-many broadcasting), when restricting agents to share only arguments that support their own posterior beliefs (non-cooperative sharing), and when salient group identities make it more likely that agents reject information from outgroup members and accept information from ingroup peers (group salience).

\section*{Discussion}
Much has already been written about the effects of homophily on opinion formation in groups, since at least social judgment theory \cite{sherif1961social}. Directly relevant to our work is modeling research on the effects concerning polarization \cite{flache2017models, deffuant2000mixing,degroot1974reaching,hegselmann2002opinion}, especially within wisdom-of-crowds settings \cite{hahn2019communication,hahn2020truth,assaad2023bayesian,hahn2024knowledge}. In the bulk of this literature, but with the notable exception of \cite{assaad2023bayesian,hahn2024knowledge}, which we come back to below, information exchange is typically modeled as a form of belief averaging, thereby abstracting away from an explicit model of the evidence that is exchanged and determines belief formation. Our modeling of information exchange aims instead to represent explicitly each individual piece of evidence that is exchanged via social interaction. This modeling choice enables novel observations that target the combined effect of homophily and the amount of information that agents are able to exchange in interaction. We model pieces of evidence as observations from Bernoulli trials with a specific quality (see SI). In doing this, we model information in a slightly more stylized manner than previous work has proposed \cite{assaad2023bayesian,hahn2024knowledge}, which is, to the best of our knowledge, the only other work attempting to model evidence explicitly. There, information is modeled using richer structures known as Bayesian networks, of which our model is a special case. The more stylized model we use allows us to represent the amount of information exchanged in a parsimonious manner, enabling us to develop a novel angle on the interplay between amount of information exchanged and homophily. 

Our findings are driven by an agent-based model methodology. They do, however, align with observations from empirical studies on the wisdom of crowds and information exchange. For example, negative epistemic effects of social influence among peers with similar ideological priors were empirically observed \cite{stein2023realtime}, although information capacity was not treated as an explicit experimental parameter. Related empirical research has focused also on the role of partisan identities, showing that group salience can decrease influence from out-group members \cite{guilbeault2018social}. While not a part of our main model, additional findings in the SI show that such measures of identity salience produce effects similar to homophily, and can complicate the epistemic progress of populations with free information flow. In showing how moderate amounts of communication can produce large epistemic gain despite high levels of homophily, our results further align with empirical findings suggesting that the wisdom of crowds is robust to homophilous partitioning \cite{becker2019wisdom}. 

A feature of opinion dynamics that is incorporated in some empirical as well as agent-based modeling research \cite{flache2017models} is negative influence, according to which contrasting ideological views can sometimes amplify polarization tendencies \cite{bail2018exposure}. Our analysis shows that even in the absence of strong polarization drivers such as negative influence, epistemic group performance declines under high homophily and information-capacity regimes.

Finally, we note that our model is circumscribed by a setting in which agents are confronted with a binary decision (cf. \cite{becker2022crowd}). While this may be applicable to many circumstances, such as forming beliefs on whether a law should be passed or not, whether a drug should be approved or disapproved, or whether a defendant is guilty or innocent, it is important to point out that other decision environments may be more complex. A natural direction for future work is to investigate if the results obtained here also apply to ordinal \cite{bikhchandani1992theory, banerjee1992quarterly} or multi-dimensional \cite{lazer2007network} decision problems.

\medskip

In general, our findings identify a core trade-off in the design of information networks between homophily and communication capacity.
When communication is homophilous, the ability to exchange large amounts of information penalizes precisely the agents who would benefit most from accessing accurate information. If such negative effects can be established in groups of ideal agents like the ones we study, caution suggests that they should remain a concern when agents' information processing abilities and cooperative tendencies are less than ideal, like in real-world information networks.

The above insight aligns with previous findings, analytical \cite{peleg2012extending,pivato2017epistemic}, simulation-based \cite{hahn2019communication,hahn2020truth}, and empirical \cite{lorenz2011social}, that highlight how communication may both improve crowd-wisdom, by disseminating high-quality information within the group, as well as damage it, by introducing too much consistency across agents' beliefs and thereby disrupting effective information aggregation.

To conclude, our results align with previous work in suggesting that groups' epistemic performance crucially depends on structural features of individuals' interactions, such as homophily. They suggest, in addition, that tradeoffs exist between epistemic performance, interaction structure, and the amount of information that individuals are able to exchange in interaction. Any implementation of the ``information wants to be free'' tenet in the design of information networks should, therefore, not ignore the structural features of interaction that constrain or facilitate information flow.

\matmethods{

\subsection*{Experimental design} We implemented our model in NetLogo, version 6.2.0 \cite{wilensky1999netlogo}. The complete replication package can be found on OSF \url{https://osf.io/vg3y2/?view_only=904baad36cb84c7db5461a1554900c02}.
The model is instantiated for a crowd of $N = 100$ agents and each experiment stopped after 1000 interactions during which no agent crossed the 0.5 posterior threshold. We conducted 1000 simulation runs per parameter combination of $h$ and $k$. Sensitivity analyses in the SI confirm that results stay similar in populations of larger sizes.

\subsection*{Environment} There are two states of the world: $A$ and $B$. We assume that $A$ is the true state and that each agent has a prior $\Pr(A) = \Pr(B) = 0.5$ as typically done in maximum-likelihood estimation settings \cite{elkind16rationalizations}. We work with a discrete time stochastic model with a finite population.

\subsection*{Initialization of agents' beliefs} At initialization (time $t = 0$), each agent $i$ is endowed with a set $S_i$ ($|S_{i,0}|<k$) of private pieces of evidence about the state of the world. These pieces of evidence are pairs $(s,q)$ where:
\begin{itemize}
    \item $s$ is a random variable which can take either value $A$ (the information supports state $A$) or $B$ (the information supports state $B$);
    \item $q$ is a parameter defining the Bernoulli distribution
    
    \begin{equation}
    q = \Pr(s \vert A) = 1 - \Pr(s \vert B) \in (0.5, 1).
    \end{equation}
    
    from which $s$ is drawn. $q$ describes the quality of the specific piece of evidence, that is, the conditional probability of receiving signal $s$ given the state of the world. We assume these quality parameters to be drawn from a beta distribution and transformed according to 
    
\begin{equation}
q \mathtt{\sim} \frac{\textrm{Beta}(2,9)}{2} + 0.5.
\end{equation}

\end{itemize}
Upon observation of these pieces of evidence $(s,q)$, agents obtain their initial belief about the state of the world via Bayes' rule: 

\begin{equation}
\B_{i,0} = Pr(A \mid S_{i,0}),
\end{equation}

that is, the belief of agent $i$ for state $A$ at time $0$. Clearly, the belief of agent $i$ for state $B$ at time $0$ is $1 - \B_{i,0}$. 

In choosing a right-skewed distribution of $Beta(2,9)$ with a relatively low average signal quality $\bar{q} = 0.59$, we ensure that a sizable minority of agents starts off with initial evidence pointing towards $B$ as the true state. Sensitivity analyses in the SI reveal that although general patterns remain unchanged, results become weaker in less right-skewed Beta distributions. Here, higher average signal quality makes it easier for agents to infer the true state of the world $A$ and harder for subpopulations to develop stable incorrect beliefs. Results are robust to a larger endowment of initial signals per agent $1 < |S| < 10$.

\subsection*{Dynamics of agents' beliefs} At each subsequent time step $t > 0$, each agent is paired with another agent with whom to interact and exchange information. This happens stochastically and agents are more likely to interact with agents similar to themselves in terms of their view of the true state of the world at the previous time $t-1$: $\B_{*,t - 1}$. The similarity score $\sigma_{i,j,t}$ between two agents $i$ and $j$ at time $t$ is calculated as 

\begin{equation}
\sigma_{i,j,t} = 1 - |\B_{j,t} - \B_{i,t}|.
\end{equation}

 The two agents are then matched with probability equal to 

\begin{equation}
\frac{\sigma_{i,j,t}^h}{\sum_{j \in J} \sigma_{i,j,t}^h}
\end{equation}

where $h \in [0,10]$ is the homophily parameter and $J = N \backslash \{i \}$ is the set of agents minus agent $i$.
Once the partner has been selected, agents exchange $k$ signals (i.e., pieces of evidence) so that 

\begin{equation}
S_{i,t+1} = S_{i,t} \cup C^t_j,
\end{equation}

where: 
\begin{itemize}
    \item $C^t_j = TOP^t_\alpha(A) \cup TOP^t_\beta(B)$, with $TOP^t_\alpha$ standing for the best (by quality) $\alpha$ signals supporting state $A$ in $j$'s observation set at time $t$ (similarly for $B$);
    \item $\alpha = \lceil \B_{j,t}(A) \cdot k \rceil$ and $\beta = \lfloor \B_{j,t}(B) \cdot k \rfloor$.
\end{itemize}
Intuitively, $j$ passes to $i$ the best (highest quality) signals she has for $A$ and $B$ proportionally to her posterior at time $t$. This information update models a form of informational `good faith' by agents, who are willing to exchange pieces of evidence that go against their current belief.

After each interaction, agents iterate over the set of newly received signals $S_{i,t} \backslash S_{i, t-1}$ to update their belief. For each new signal supporting $A$, they use Bayes' rule to obtain: 

\begin{equation}
\B_{i,t+1} =  \frac{q \cdot \B_{i,t}}{q \cdot \B_{i,t}  + ((1 - q) \cdot (1 - \B_{i,t}))},
\end{equation}

where $q$ is the quality of the signal received, giving thus also the belief for state $B$: $1 - \B_{i,t+1}$.
As a result of this operationalization, as populations polarize, similarity scores $\sigma$ increase for like-minded agents and decrease for agents with opposing beliefs. Consequently, receivers become more likely to select interaction partners who share their views. At the same time, senders gain confidence in their beliefs and selectively share arguments more in line with their posterior. These dynamics generate a feedback loop: rising polarization increases the likelihood of interactions with similar others, who in turn provide information that further reinforces existing views.

We computed the {\em realized information gain} as: 
\begin{equation}
    \rho_i = \frac{\alpha_i}{\phi_i},
\end{equation}
where $\phi_i = 1 - B_i$ is the difference between the full-information posterior $1$ and $i$'s initial belief. Analysis at the agent-level comprised  multilevel models with agents ($N = 100$) nested within runs ($N = 70,000$), with individual-level epistemic gain ($\rho_i$) as the dependent variable. Analysis at the group-level and the population-level were performed through linear regressions.

}

\showmatmethods{} % Display the Materials and Methods section

\acknow{
We would like to thank the Lorentz Centre, the organisers - Dr Francesca Giardini, Dr Jens Madsen, Prof Jos Hornikx - and the participants of the \href{https://www.lorentzcenter.nl/modelling-social-complexity-in-argumentation.html}{"Computational modelling of social complexity in argumentation"} for giving us the opportunity to meet each other and develop this project.

Davide Grossi acknowledges support from the \href{https://hybrid-intelligence-centre.nl}{Hybrid Intelligence Center}, a 10-year program funded by the Dutch
Ministry of Education, Culture and Science through the Netherlands Organisation for Scientific Research (NWO) and by the European Union under the Horizon Europe project \href{https://perycles-project.eu/}{Perycles} (Participatory Democracy that Scales). Views and opinions expressed are however those of the author only and do not necessarily reflect those of the European Union or the European Research Executive Agency (REA). Neither the European Union nor the granting authority can be held responsible for them.

\smallskip
\begin{center}
\includegraphics[width=0.3\textwidth]{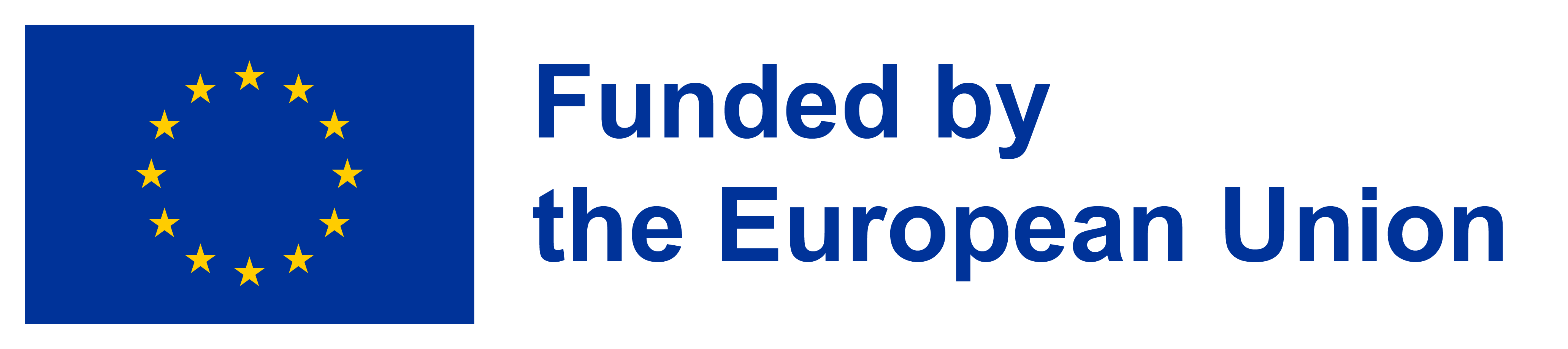}
\end{center}

}

\showacknow{} % Display the acknowledgments section

% Bibliography
%\bibliography{references}

\end{document}